# An Efficient Approach for Energy Conservation in Cloud Computing Environment

Sohan Kumar Pande[1], Sanjaya Kumar Panda[2], *Member*, *IEI*, and Preeti Ranjan Sahu[3]
[1,3]*Veer Surendra Sai University of Technology*, *Burla – 768018*, *Odisha*, *India*
[2]*National Institute of Technology*, *Warangal – 506004*, *India*
{[1]ersohanpande, [2]sanjayauce, [3]preetiranjan.sahu}@gmail.com

*Abstract*—Recent trends of technology have explored a numerous applications of cloud services, which require a significant amount of energy. In the present scenario, most of the energy sources are limited and have a greenhouse effect on the environment. Therefore, it is the need of the hour that the energy consumed by the cloud service providers must be reduced and it is a great challenge to the research community to develop energy-efficient algorithms. To design the same, some researchers tried to maximize the average resource utilization, whereas some researchers tried to minimize the makespan. However, they have not considered different types of resources that are present in the physical machines. In this paper, we propose a task scheduling algorithm, which tries to improve utilization of resources (like CPU, disk, I/O) explicitly, which in turn increases the utilization of active resources. For this, the proposed algorithm uses a fitness value, which is a function of CPU, disk and I/O utilization, and processing time of the task. To demonstrate the performance of the proposed algorithm, extensive simulations are performed on both proposed algorithm and existing algorithm MaxUtil using synthetic datasets. From the simulation results, it can be observed that the proposed algorithm is a better energy-efficient algorithm and consumes less energy than the MaxUtil algorithm.

## I. INTRODUCTION

In the era of modern technology, cloud computing has gained huge popularity for providing services like infrastructure as a service (IaaS), platform as a service (PaaS) and software as a service (SaaS) [1, 2]. People who face a shortage of resources to complete their required computation work, they get benefit from the cloud services and pay a marginal amount to use them [1, 3]. The services are provided by the cloud service providers (CSPs) [1-3]. CSP handles all the tasks submitted by the users by creating virtual machines (VMs) on the physical machines (PMs) or using the resources present at the datacenter [1-4]. It is required that the resources present at the datacenter must be handled efficiently by the CSP, so that maximum tasks can be executed in minimum time. It is noteworthy to mention that minimizing the processing time and maximizing the average resource utilization does not guarantee the minimum consumption of energy.

In the present situation, most of the industries and service providers are relying on the non-renewable energy [5]. The sources of the non-renewable energy are limited in nature, which will exhaust one day and these resources have an adverse effect on the environment [5]. Therefore, the need of the hour is that the consumption of energy for different computational work and services must be minimized. The incurring cost of providing services will also reduce, if there will be efficient energy management. Most of the recent works are either focused on the minimizing the processing time or maximizing the average resource utilization [3, 4]. Some of the works also focus on efficient energy utilization by considering only the resource utilization and ignoring the processing time [4-6]. Moreover, in the majority of works, individual resource utilizations (like CPU, disk and I/O) are not properly addressed. Therefore, in this paper, we consider both the processing time and the average resource utilization of CPU, disk and I/O to achieve efficient energy utilization and propose a multi-criteria based energy-efficient task scheduling (MCEETS) algorithm.

The proposed algorithm is a two-phase algorithm, where the first phase deals with the ordering of the tasks arrived at the same time for the execution purpose and the second phase deals with the assignment of the task to a VM in which the task can be accommodated with respect to the availability of the different types of resources. While ordering the tasks in the first phase, the fitness value for each task is calculated. The fitness value depends on both the processing time and the resource utilization. In this paper, we consider mainly three types of resource utilization and they are (1) CPU utilization (b) disk utilization and (c) I/O utilization. The final ordering of the tasks is prepared by organizing the tasks in the ascending order of their fitness value. In the second phase, one task is selected from the list, which is prepared in the first phase and that task is assigned to a suitable VM, where the requirement of the task is fulfilled. If there are more than one VMs suitable for that task, then the task is assigned to a VM, where the maximum resource utilization is achieved. As a result, energy consumed for executing the tasks is minimized. Therefore, the main contributions of this paper are as follows. (1) MCEETS considers both the processing time and the average resource utilization to achieve energy-efficient scheduling. (2) The proposed algorithm considers different types of resource utilization (like CPU, disk and I/O). (3) The simulation results of MCEETS are compared with the existing algorithm MaxUtil using synthetic datasets.

The rest of the paper is structured as follows. Section II contains the overview of the recent research works on scheduling algorithms to achieve better energy efficiency and resource utilization. The proposed cloud model and the details of the problem is discussed in the section III. MCEETS is explained with pseudo code in section IV. Section V discuses about the simulation results and a comparison with the existing algorithm MaxUtil. Section VI contains the concluding remarks of the presented MCEETS algorithm.

## II. RELATED WORK

Cloud computing is an evolving paradigm in the field of IT, which gives services to the users facing shortage of resources and the service providers are compensated by the users [1-5]. To provide different types of cloud services, datacenters of the CSPs require a huge amount of energy. For example, the electricity used by the Google datacenter is approximately same as the electricity consumed by a big city. Most of the energy sources are non-renewable and have an adverse effect on the environment, which is an ultimate cause of greenhouse effect [5]. Therefore, it is very much required by the datacenters to adopt efficient scheduling algorithms, which can minimize the energy consumption without compromising in QoS promised to the users.

Many research works have shown that energy consumption of the datacenter is linearly dependent on the average resource utilization. Lee et al. [4] have proposed an energy-conscious task consolidation algorithm, which explicitly addresses the issue of energy consumed by the ideal resources and active resources. Their proposed algorithms assign the tasks to the VMs, such that minimum energy is consumed. Panda et al. [3] have proposed three algorithms, MCC, MEMAX and CMMN, which give better makespan and maximum resource utilization. Minimizing makespan directly reduces the total energy consumed by the datacenters. Some research works show that VM in ideal state or VM with low load also consumes a significant amount of energy. Therefore, researchers have proposed algorithms, which transfer tasks from overloaded VMs to under loaded VMs and shares loads among the VMs [2, 6-10].

While some researchers try to balance the loads between the VMs to achieve energy efficiency, some studies have shown that VMs, with utilization more than 70%, also consume a high amount of energy [6]. Therefore, Hsu et al. [6] have proposed task consolidation techniques, which define a threshold value and when the utilization of the VM crosses that threshold value, some of the load from the overloaded VM is migrated to some under loaded VM. The authors have also considered utilization and network transmission for the task consolidation purpose. Dabbagh et al. [2] have proposed a resource management framework, which predicts accurate number of VMs needed with required CPU and memory resources. Then this framework accurately estimates PMs needed and puts other PMs into sleep mode. Beloglazov et al. [7] have found different challenges in cloud architecture to gain better energy utilization. These research works have not considered different types of resources like CPU, disk and I/O, while assigning loads to the VMs. However, considering individual resource can significantly increase the average resource utilization, which will ultimately give better energy utilization. In this research work, we consider different types of resources along with the processing time to select a better VM.

## III. CLOUD MODEL AND PROBLEM STATEMENT

### A. Cloud Model

We consider various cloud services, which is provided to the users on-demand basis. In this model, users submit their tasks (user requests) to the CSP and pay for the same. To compute the tasks submitted by the users, VMs are created on the PM in the datacenter of the CSP, so they can fulfil the demand of the users. All these VMs are managed by the datacenter in terms of computational capacity and it is assumed that all the VMs are homogeneous irrespective of the PM on which it is hosted. When the CSP receives the task, it forwards the task to the datacenter and the datacenter in turn assigns the task to one of the created VM. For executing the task on the VM, energy is consumed by the respective PM on which the VM is hosted. The energy consumed at a particular time by the PM is directly proportional to the current resource utilization of the VM. Here, we consider three types of resource utilization, namely CPU, disk and I/O utilization. One VM can execute more than one task at a time as per the availability of the resources. At any point of time $t$, utilization of CPU ($UC$) offered by a VM $i$ is the total of the utilization of the resource by all the tasks executing at that time on that respective VM. Mathematically, it can be written as

$$UC_i^t = \sum_{j=1}^{n} UC_j^t \times assign_j^t \qquad (1)$$

where $assign_j^t = \begin{cases} 1 & \text{if task } j \text{ is assigned at time } t \\ 0 & \text{Otherwise} \end{cases}$

and $n$ is the number of tasks executing at time $t$. Similarly, we calculate utilization of the disk as follows.

$$UD_i^t = \sum_{j=1}^{n} UD_j^t \times assign_j^t \qquad (2)$$

We calculate utilization of the I/O as follows.

$$UI_i^t = \sum_{j=1}^{n} UI_j^t \times assign_j^t \qquad (3)$$

The total utilization of the VM $i$ can be represented as the average of the CPU, disk and I/O utilization at time $t$, it can be mathematically represented as

$$UV_i^t = \frac{(UC_i^t + UD_i^t + UI_i^t)}{3} \qquad (4)$$

The average utilization of the VM $i$ with a time period $T$ will be $UV_i = \frac{1}{T}\sum_{t=1}^{T} UV_i^t$. The energy consumed by the VM $i$ can be calculated as $E_i = (P_{max} - P_{min}) \times UV_i + P_{min}$. The Total energy consumed by the PM is the total energy consumed by all VMs hosted on that PM, that can be expressed as $E = \sum_{i=1}^{m} E_i$. Here, $m$ is the number of VMs hosted on the PM.

### B. Problem Statement

Let us consider that $n$ number of tasks is submitted by the users to the CSP and the datacenter of the CSP has to assign these tasks to $m$ number of VMs hosted on the PM. Each task $T_j$, $1 \leq j \leq n$ is represented in 7-tuple as follows, {*TID*, *TAT*, *TPT*, *TFT*, *TCU*, *TDU*, *TIU*}. These tuples denote task id, task arrival time, task processing time, task finish time, task CPU utilization, task disk utilization and task I/O utilization, respectively. We can say *TFT* = *TAT* + *TPT*. Here, the problem is to assign the tasks to the available VMs in such a way that total energy consumption by all the VMs is minimized. This assignment problem is subject to the following constraints. (1) At a time *t*, a task *j* can only be assigned to one VM *i* and the *TAT* ≤ *t*. (2) Task *j* can only be assigned to VM *i* at time *t* only if $CU_i^t + TCU^t \leq 100$ and $DU_i^t + TDU^t \leq 100$ and $IU_i^t + TIU^t \leq 100$.

IV. PROPOSED ALGORITHM

The proposed consolidation algorithm comprises of two phases. The first phase of the algorithm finds the ordering of the tasks for assigning them to the VMs if more than one task arrives at the same time. The ordering of the tasks is decided based on the fitness value of the tasks. The fitness value of the task considers CPU utilization, disk utilization, I/O utilization and processing time. Mathematically, it can be represented as follows.

$$F = \lambda \times NTPT + (1 - \lambda) \times (NCU + NDU + NIU), \text{ here } 0 \leq \lambda \leq 1 \quad (5)$$

After finding the fitness values of the tasks, they are arranged in the ascending order of their fitness value. In the second phase, the tasks are assigned to the VMs, such that resources (CPU, disk and I/O) utilization will be minimized, which ultimately consume less energy for computation.

TABLE I. NOTATIONS AND THEIR DEFINITION

| Notation | Definition |
|---|---|
| Q | Global queue |
| N | Number of tasks |
| M | Number of VMs |
| T | Maximum time to compute all the task |
| TC(j, 1), TC(j, 2),…, TC(j, 6) | Task characteristics *TAT*, *TPT*, *TFT*, *TCU*, *TDU*, *TIU*, respectively |
| NOTAT | Number of unique arrival time |
| Maxtime | Maximum of all the arrival time |
| NOT(t) | Total number of tasks arrived at time *t* |
| TASK_LIST | List of the tasks arrive at time *t* |
| Ct | Current task to be executed |
| Novm | Number of VMs available to execute ct |
| dest_vm | Destination VM |
| UV(i) | Average resource utilization of VM *i* |
| TE | Total energy |

| Algorithm: MCEETS |
|---|
| **Input:** 2-D matrix: *TC*, *n*: number of tasks, *m*: number of VMs |
| **Output:** *UV*, *TE* |
| 1.  **while** *Q* ≠ NULL |
| 2.       *maxtime* = *TC*(1, 1) |
| 3.       **for** *i* = 1:*n* |
| 4.           *NOTAT*(*TC*(*j*, 1)) = *NOTAT*(*TC*(*j*, 1)) + 1 |
| 5.           **if** *maxtime* < *TC*(*j*, 1) |
| 6.               *maxtime* = *TC*(*j*, 1) |
| 7.           **endif** |
| 8.       **endfor** |
| 9.       **for** *t* = 1:*maxtime* |
| 10.          **if** *NOTAT*(*t*) ≠ 0 |
| 11.              **for** *j* = 1:*n* |
| 12.                  **if** *TC*(*j*, 1) = *t* |
| 13.                      *NOT* = *NOT* + 1 |
| 14.                      *TASK_LIST*(*NOT*) = *j* |
| 15.                  **endif** |
| 16.              **endfor** |
| 17.              *max_pt* = *TC*(1, 2) |
| 18.              **for** *j* = 1:*NOTAT*(*t*) |
| 19.                  **if** *max_pt* < *TC*(*TASK_LIST*(*j*, 2)) |
| 20.                      *max_pt* = *TC*(*TASK_LIST*(*j*, 2)) |
| 21.                  **endif** |
| 22.              **endfor** |
| 23.              **for** *j* = 1:*NOTAT*(*t*) |
| 24.                  *NPT*(*j*) = *TC*(*TASK_LIST*(*j*), 2) / *max_pt* |
| 25.              **endfor** |
| 26.              *NCU*, *NDU* and *NIU* values are found in the same way as *NPT* is found (Lines 17-25) |
| 27.              **for** *j* = 1:*NOTAT*(*t*) |
| 28.                  *fitness*(*j*) = 0.5 × *NPT*(*j*) + 0.5 × (*NCU*(*j*) + *NDU*(*j*) + *NIU*(*j*)) |
| 29.              **endfor** |
| 30.              **for** *j* = 1:*TOTAT*(*t*) – 1 |
| 31.                  **for** *j'* = *j*:*TOTAT*(*t*) |
| 32.                      **if** *fitness*(*j*) > *fitness*(*j'*) |
| 33.                          *temp* = *fitness*(*j*) |
| 34.                          *fitness*(*j*) = *fitness*(*j'*) |
| 35.                          *fitness*(*j'*) = *temp* |
| 36.                          *temp1* = *TASK_LIST*(*j*) |
| 37.                          *TASK_LIST*(*j*) = *TASK_LIST*(*j'*) |
| 38.                          *TASK_LIST*(*j*) = *temp1* |
| 39.                      **endif** |
| 40.                  **endfor** |
| 41.                  Call *ASSIGN_TASK_VM*(*TASK_LIST*, *n*, *m*) |
| 42.              **endfor** |
| 43.          **endif** |
| 44.          Call *CALCULATE_UV_E*(*TC*, *assign_vm*, *m*, *n*) |
| 45. **endwhile** |

Fig. 1. Pseudo code for MCEETS.

The pseudo code for the proposed algorithm is presented in Fig 1. For this, notations and their definition are given in the Table I. Let's us now explain the pseudo code for the proposed algorithm MCEETS with the help of an example. Suppose there are 10 number of tasks to execute on the VMs created by the datacenter of the CSP. All the VMs are homogenous in terms of computing capacity and resource availability. The detailed features of the tasks are presented in the Table II.

TABLE II. DETAILED FEATURES OF TASKS

| TID | TAT | TPT | TFT | TCU | TDU | TIU |
|---|---|---|---|---|---|---|
| $T_1$ | 1 | 25 | 26 | 30 | 22 | 30 |
| $T_2$ | 1 | 29 | 30 | 31 | 21 | 31 |
| $T_3$ | 1 | 23 | 24 | 32 | 27 | 32 |
| $T_4$ | 2 | 32 | 34 | 24 | 25 | 25 |
| $T_5$ | 2 | 24 | 26 | 30 | 23 | 31 |
| $T_6$ | 2 | 28 | 30 | 30 | 31 | 31 |
| $T_7$ | 2 | 31 | 33 | 22 | 24 | 22 |
| $T_8$ | 3 | 34 | 37 | 21 | 24 | 21 |
| $T_9$ | 3 | 35 | 38 | 27 | 30 | 27 |
| $T_{10}$ | 3 | 28 | 31 | 35 | 30 | 25 |

After submission of the tasks to the datacenter, tasks are put in queue *Q* (Fig 1, Line 1). Then number of tasks arrived at different time is calculated (Lines 2-8). After this step, task list is prepared, which are arrived at the same time (Lines 11-16). In our example, ten tasks ($T_1$ to $T_{10}$) are arrived at different time. For instance, at time $t = 1$, $T_1$, $T_2$ and $T_3$ have arrived. Similarly, at time $t = 2$, $T_4$, $T_5$, $T_6$ and $T_7$ and at time $t = 3$, $T_8$, $T_9$ and $T_{10}$ have arrived. After finding the tasks which have arrived at same time, we have to find the order in which they will be assigned to the VMs. To find the ordering of the tasks, fitness value is calculated. To find out the fitness value of each task, first the normalized processing time (*NPT*) of each task is calculated (Lines 17-25). Similarly, we calculate *NCU*, *NDU* and *NIU* (Lines 26). Then fitness value is calculated (Lines 27-29). In our example, at time $t = 1$, $T_1$, $T_2$ and $T_3$ have arrived. For $T_1$, fitness value is calculated as fitness = 0.5 × (25/29) + 0.5 × [ (30/32) + (22/27) + (30/32)] = 1.775. Similarly, the fitness value of $T_2$ and $T_3$ are found to be 1.857 and 1.896. Then the tasks are arranged in the ascending order of their fitness value (Lines 30-40). In this example, at time $t = 1$, the ordering of the task is $T_1$, $T_2$ and $T_3$ by following the fitness value. In the similar fashion, rest of the tasks are processed in the first phase of MCEETS.

In the second phase of MCEETS, tasks are assigned to the VMs in such a way that the total energy consumed to compute all the tasks is minimized. In the second phase, the number of iteration to be performed is found out. The total number of iteration is the maximum of the *TFT* of all the tasks present in the *TASK_LIST*. In our example, the total number of iteration will be 30 as it is the maximum of the *TFT* of $T_1$, $T_2$ and $T_3$. Then from the current list of the tasks, task present in the first position is selected. We can call it current task (*ct*). Now, we have to find out the available VMs to which the task can be accommodated. One task can only be assigned to a VM, if the summation of the used resource of the VM by previously assigned tasks and the percentage of the resource required by the *ct* is less than 100% at time *t*. The VMs, who satisfy the above mentioned condition, are found out for the assignment purpose. If no VM is found out to suitable for the task, then a new VM is awaked from the sleep state. If more than one VMs are suitable for that task, then we have to find the VM on which the total resource utilization will be minimum. For this, we have to find the CPU utilization of all the suitable VMs in which task *ct* can be assigned. Then the normalized value of the CPU utilization (*norm_est_uc*) of the VMs is determined. In the similar fashion, *norm_est_ud* and *norm_est_ui* are calculated for each suitable VM. Then total normalized utilization value for each VM is found out. After finding out the total normalized utilization value, the VM having the maximum value is found out and the task *ct* is assigned. After the task is assigned to the VM, CPU utilization, disk utilization and I/O utilization of the respective VM is updated. Then total energy consumed by all the VMs are calculated. These procedures are repeated until all the task are assigned to the VMs. Note that the procedures are not shown in this paper due to space constraint. In our example, the tasks are assigned to the VMs as shown in the Gantt chart (Fig. 4 to Fig. 12).

| $VM_1$ | 1~30 | 30~61 | 61~93 | 93~100 |
|---|---|---|---|---|
| 1~24 | $T_1$ | $T_2$ | $T_3$ | * |
| 24~26 | | | * | * |
| 26~30 | * | | * | * |
| 30~37 | * | * | * | * |

Fig. 2. *UC* of $VM_1$ for MCEETS.

| $VM_2$ | 1~22 | 22~46 | 46~76 | 76~97 | 97~100 |
|---|---|---|---|---|---|
| 2~3 | | | | * | * |
| 3~26 | $T_7$ | $T_4$ | $T_5$ | | * |
| 26~33 | | | | $T_8$ | * |
| 33~34 | * | | * | | * |
| 34~37 | * | * | * | | * |
| 37~38 | * | * | * | * | * |

Fig. 3. *UC* of $VM_2$ for MCEETS.

| $VM_3$ | 1~30 | 30~65 | 65~92 | 92~100 |
|---|---|---|---|---|
| 2~3 | $T_6$ | * | * | * |
| 3~30 | | $T_{10}$ | $T_9$ | * |
| 30~31 | * | | | * |
| 31~38 | * | * | | * |

Fig. 4. *UC* of $VM_3$ for MCEETS.

| $VM_1$ | 1~22 | 22~43 | 43~70 | 70~100 |
|---|---|---|---|---|
| 1~24 | $T_1$ | $T_2$ | $T_3$ | * |
| 24~26 | | | * | * |
| 26~30 | * | | * | * |
| 30~37 | * | * | * | * |

Fig. 5. *UD* of $VM_1$ for MCEETS.

| $VM_2$ | 1~24 | 24~49 | 49~72 | 72~96 | 96~100 |
|---|---|---|---|---|---|
| 2~3 | | | | * | * |
| 3~26 | $T_7$ | $T_4$ | $T_5$ | | * |
| 26~33 | | | * | $T_8$ | * |
| 33~34 | * | | * | | * |
| 34~37 | * | * | * | | * |
| 37~38 | * | * | * | * | * |

Fig. 6. *UD* of $VM_2$ for MCEETS.

| $VM_3$ | 1~31 | 31~61 | 61~91 | 91~100 |
|---|---|---|---|---|
| 2~3 | $T_6$ | * | * | * |
| 3~30 | | $T_{10}$ | $T_9$ | * |
| 30~31 | * | | | * |
| 31~38 | * | * | | * |

Fig. 7. *UD* of $VM_3$ for MCEETS.

| $VM_1$ | 1~30 | 30~61 | 61~93 | 93~100 |
|---|---|---|---|---|
| 1~24 | $T_1$ | $T_2$ | $T_3$ | * |
| 24~26 | | | * | * |
| 26~30 | * | | * | * |
| 30~37 | * | * | * | * |

Fig. 8. *UI* of $VM_1$ for MCEETS.

| $VM_2$ | 1~22 | 22~47 | 47~68 | 68~89 | 89~100 |
|---|---|---|---|---|---|
| 2~3 | | | | * | * |
| 3~26 | $T_7$ | $T_4$ | $T_5$ | | * |
| 26~33 | | | * | $T_8$ | * |
| 33~34 | * | | * | | * |
| 34~37 | * | * | * | | * |
| 37~39 | * | * | * | * | * |

Fig. 9. *UI* of $VM_2$ for MCEETS.

| $VM_3$ | 1~31 | 31~56 | 56~83 | 83~100 |
|---|---|---|---|---|
| 2~3 | $T_6$ | * | * | * |
| 3~30 | | $T_{10}$ | $T_9$ | * |
| 30~31 | * | | | * |
| 31~38 | * | * | | * |

Fig. 10. *UI* of $VM_3$ for MCEETS.

The average CPU utilization of $VM_1$ is calculated as $UC_1 = (30 \times 25 + 31 \times 29 + 23 \times 32)/29 = 82.24$. Like this, $UC_2$, $UC_3$, $UD_1$, $UD_2$, $UD_3$, $UI_1$, $UI_2$ and $UI_3$ are calculated. These values are shown in the Table III.

TABLE III. RESOURCE UTILIZATION OF VM FOR MCEETS

| VM | UC | UD | UI |
|---|---|---|---|
| $VM_1$ | 82.24 | 61.30 | 82.24 |
| $VM_2$ | 88.40 | 83.20 | 77.10 |
| $VM_3$ | 76.80 | 76.60 | 69.80 |

We have assumed $pmax$ and $pmin$ as 30 and 20. The energy consumed by $VM_1$ is calculated as $E_1 = (30 - 20) \times ((82.24 + 61.30 + 82.24)/3) + 20 = 772.6$. Similarly, energy consumed by $VM_2$ and $VM_3$ is found to be 842.9 and 764, respectively. Therefore, a total of 2379.5 energy units is used to compute all the tasks.

We have compared the proposed algorithm with the existing algorithm MaxUtil, which assigns the tasks in chronological order to the VMs. The Gantt charts for the existing algorithm is not shown due to space constraint. The average resource utilization by the VMs for the existing algorithm is given in the Table IV.

TABLE IV. RESOURCE UTILIZATION OF VM FOR MAXUTIL

| VM | UC | UD | UI |
|---|---|---|---|
| $VM_1$ | 84.44 | 63.24 | 84.44 |
| $VM_2$ | 73.00 | 69.66 | 75.70 |
| $VM_3$ | 65.10 | 72.64 | 65.10 |
| $VM_4$ | 35.00 | 30.00 | 25.00 |

The energy consumed for existing algorithm MaxUtil is 793.7, 747.86, 696.13 and 320 for $VM_1$, $VM_2$, $VM_3$ and $VM_4$, respectively. Therefore, the total energy consumed by MaxUtil algorithm to compute all the tasks is 2557.69 energy units, whereas the proposed algorithm takes 2379.5 energy units. In this example, MCEETS consumes 6.96% less energy than the existing algorithm MaxUtil. In the proposed algorithm, a total of 3 number of VMs is used, whereas a total of 4 numbers of VMs is used in the existing algorithm MaxUtil. The detailed comparison between the proposed and existing algorithms is given in Table V.

TABLE V. COMPARISON OF PERFORMANCE METRICS

| Performance Metrics | MCEETS | MaxUtil |
|---|---|---|
| TE | 2379.5 | 2557.69 |
| UV | [75.26, 82.29, 74.4] | [77.37, 72.28, 67.61, 30] |
| VMs Used | 3 | 4 |

TABLE VI. PARAMETER AND THEIR VALUES

| Parameter | Values |
|---|---|
| # of Tasks | 100, 500, 1000, 5000, 10000 |
| # of VMs | 20, 100, 200, 1000, 2000 |
| Instances | $i1, i2, i3, i4, i5, i6, i7, i8, i9, i10$ |
| TAT, TPT | [1 ~ 50], [20 ~ 100] |
| TFT, TCU | [1 ~ 150], [5 ~ 20] |
| TDU, TIU | [5 ~ 20], [5 ~ 20] |

TABLE VII. COMPARISON OF TE FOR MCEETS AND MAXUTIL

| #Tasks × #VMs | Instance | MCEETS | MaxUtil | Instance | MCEETS | MaxUtil |
|---|---|---|---|---|---|---|
| 100 × 20 | i1 | 8110 | 8218 | i6 | 7424 | 7563 |
| | i2 | 8110 | 8219 | i7 | 8032 | 8761 |
| | i3 | 8065 | 8093 | i8 | 7637 | 8010 |
| | i4 | 7935 | 7996 | i9 | 8077 | 8652 |
| | i5 | 8267 | 8383 | i10 | 7944 | 7979 |
| 500 × 100 | i1 | 39898 | 43974 | i6 | 40123 | 40668 |
| | i2 | 38549 | 38576 | i7 | 39396 | 39746 |
| | i3 | 40304 | 41564 | i8 | 38651 | 38772 |
| | i4 | 41511 | 42758 | i9 | 39285 | 39485 |
| | i5 | 40375 | 42169 | i10 | 38569 | 39764 |
| 1000 × 200 | i1 | 81107 | 81479 | i6 | 78441 | 78652 |
| | i2 | 79695 | 80093 | i7 | 80117 | 85764 |
| | i3 | 78704 | 78969 | i8 | 78463 | 82345 |
| | i4 | 79151 | 79756 | i9 | 80083 | 87562 |
| | i5 | 78136 | 78374 | i10 | 79988 | 86357 |
| 5000 × 1000 | i1 | 397000 | 397112 | i6 | 404073 | 446539 |
| | i2 | 402193 | 417643 | i7 | 400959 | 417324 |
| | i3 | 400639 | 400901 | i8 | 396002 | 397970 |
| | i4 | 402662 | 423562 | i9 | 394798 | 395069 |
| | i5 | 396852 | 399201 | i10 | 398811 | 399956 |
| 10000 × 2000 | i1 | 799748 | 805342 | i6 | 805021 | 817324 |
| | i2 | 797279 | 798345 | i7 | 795623 | 816432 |
| | i3 | 794357 | 805364 | i8 | 799176 | 825194 |
| | i4 | 802634 | 817563 | i9 | 785844 | 797828 |
| | i5 | 802340 | 823456 | i10 | 785844 | 796801 |

V. SIMULATION RESULTS

For better comparison of the proposed and existing algorithms, we conducted extensive simulation using MATLAB R2017a version 9.2.0.538062. The platform used for this simulation is Intel(R) Core (TM) i5-4210U CPU @ 2.70 GHz 2.70 GHz CPU and 8 GB RAM running on Windows 10. We evaluated both the proposed and existing algorithms using the synthetic datasets which is generated using Monte Carlo simulation method on MATLAB. The structure of the datasets is $xx\_yy\_iz$, where $xx$ represents the number of tasks to be executed, $yy$ represents the numbers of VMs and $iz$ represents the instance. We considered a large number of tasks, such as 100, 500, 1000, 5000 and 10000, and a large number of VMs as 20, 100, 200, 1000 and 2000, respectively. In each dataset, we considered 10 instances, namely $i1$ to $i10$ for simulation. The details of the parameters and their respective values of the generated datasets are presented in Table VI. In the created environment, all the datasets are simulated using both the MCEETS and MaxUtil algorithms and the results are shown in Table VII. It can be clearly observed that the MCEETS algorithm gives better performance than the MaxUtil algorithm.

VI. CONCLUSION

In this paper, we proposed a novel task consolidation algorithm MCEETS, which is an energy-efficient algorithm. In MCEETS, different types of resources, such as CPU, disk and I/O are considered to improve the average resource utilization. For the ordering of the tasks, fitness value of the tasks is calculated, which is dependent on both the processing time of the task and the resource utilization, which minimizes the energy consumption. To demonstrate the comparison of the proposed and existing algorithms, we simulated both the algorithms with synthetic datasets, where each datasets is comprised of ten different instances. From the simulated results, it can be concluded that the MCEETS performs better than the MaxUtil.

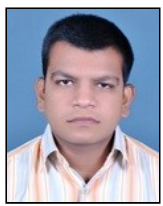

**Sohan Kumar Pande** is working as an Assistant Professor in the Department of CSE at SIT, Sambalpur, India and pursuing Ph. D. degree in the Department of CSE at VSSUT, Burla, India. He received M. Tech. degree from SUIIT, Burla, India and B. Tech. degree from CVRCE, Bhubaneswar, India in CSE. His research interests include Cloud Computing, Distributed Computing and Load Balancing. He has published more than 5 papers in reputed international journals and conferences, and acted as reviewers in many international journals, including IEEE, Elsevier Science, Springer and many more.

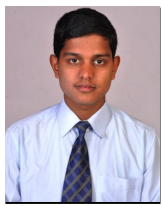

**Sanjaya Kumar Panda** is working as an Assistant Professor in the Department of CSE at NIT, Warangal, India. He received Ph. D. degree from IIT (ISM), Dhanbad, India, M. Tech. degree from NIT, Rourkela, India and B. Tech. degree from VSSUT, Burla, India in CSE. His research interests include recommender systems, cloud computing, fault tolerance and load balancing. Dr. Panda received two silver medal awards for best graduate and best post-graduate in CSE. He also received IEEE brand ambassador award for the year of 2018 and 2019, SGSITS national award for the best research work by young teachers of engineering college for the year of 2017, faculty with maximum publishing in CSI publications award, young IT professional award (2017 and 2016), young scientist award, CSI paper presenter award at international conference and CSI distinguished speaker award. Dr. Panda is an associate member of IEI, a life member of ISTE and CSI, and a member of IEEE, IAENG, IACSIT, UACEE and ACEEE. Dr. Panda has published more than 70 papers in reputed journals and conferences, and acted as reviewers in IEEE transactions on systems, man and cybernetics, IEEE/CAA journal of automatic sinica, IEEE sensors letters, journal of parallel and distributed computing, Elsevier, applied soft computing, Elsevier, the journal of supercomputing, Springer and many more.

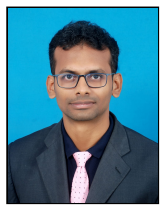

**Preeti Ranjan Sahu** is working as an Assistant Professor in the Department of EE at NIST, Berhampur, India and pursuing Ph. D. degree in the Department of EE at VSSUT, Burla, India. He received M. Tech. degree from VSSUT, Burla, India and B. Tech. degree from BPUT, Rourkela, India. His research interests include energy conservation techniques, flexible AC transmission systems, power system stability and optimization techniques. He published more than 10 papers in reputed journals and conferences, and acted as reviewers in many international journals and conferences.